An Assurance Framework for Independent Co-assurance of Safety and Security

Nikita Johnson, MEng; University of York; York, UK

Tim Kelly, DPhil; University of York; York, UK

Abstract

Integrated safety and security assurance for complex systems is difficult for many technical and socio-technical reasons such as mismatched processes, inadequate information, differing use of language and philosophies, *etc.*. Many co-assurance techniques rely on disregarding some of these challenges in order to present a unified methodology. Even with this simplification, no methodology has been widely adopted primarily because this approach is unrealistic when met with the complexity of real-world system development.

This paper presents an alternate approach by providing a Safety-Security Assurance Framework (SSAF) based on a core set of assurance principles. This is done so that safety and security can be co-assured independently, as opposed to unified co-assurance which has been shown to have significant drawbacks. This also allows for separate processes and expertise from practitioners in each domain. With this structure, the focus is shifted from simplified unification to integration through exchanging the correct information at the right time using synchronisation activities.

## Introduction

Large technological systems produce new capabilities that allow innovative solutions to social, engineering and environmental problems. This trend is especially important in the safety-critical systems (SCS) domain where we simultaneously aim to do more with the systems whilst reducing the harm they might cause. Although there are many advantages to using these new capabilities, there is also an increased risk associated with this kind of innovation. The lack of previous data and the poor understanding we have of complex system interactions mean that there is an exponentially large number of risks to evaluate and a high level of uncertainty. However, SCS still need to be assured against risk and, in many cases, certified before use.

There has been work done to create ontologies and technical mappings between the safety and security (Firesmith, 2003), yet this is still far removed from providing us with a basis for integrating the two attributes and producing a workable solution. Part of the problem is the heterogeneity of safety and security philosophies, principles and standards. They are so different that it becomes difficult to establish common ground for communication of assurance risk. It is in this context that we consider whether a principled approach relying on assurance cases can provide the necessary structure for bringing the two domains together.

In this paper we discuss the technical and socio-technical aspects of the safety-security challenge. A concise outline a candidate solution to these challenges is proposed: the Safety-Security Assurance Framework (SSAF). Projected outcomes of the framework and next steps are also discussed.

## The Safety-Security Challenge

**Technical Aspects of the Challenge**

The technical challenge describes the difficulties of integrating the two attributes in practical terms. Traditional methods for safety assurance and security assurance have been predominantly independent with little interaction through the system development life-cycle (SDLC). This is problematic because there can be little confidence in the safety argument of a system if security considerations have not been made (Bloomfield, Netkachova, & Stroud, 2013). In addition, the siloed approach leads to a conflict of concerns, and the impact on the system is detected much later on in the process when change is more costly. To ameliorate this negative effect several analysis methods and standards have been developed. The following subsections describe a subset of state-of-the-art solutions that have been applied to industrial case studies:

**Analysis Methods.** Identifying both safety and security risks during the SDLC is difficult as there may be insufficient information to perform traditional risk analyses. These methods describe approaches to integrating safety and security processes:

*Security-Aware STPA.* The Systems Theoretic Process Analysis (STPA) (Leveson, 2004) is extensively used in industry. STPA-Sec (Young & Leveson, 2013), and STPA-SafeSec (Friedberg et al., 2017) extend the STPA safety process to include security considerations. A key advantage of utilising this process is that practitioners are already familiar with it and can immediately include additional steps to account for security risk. However, when applied to a real-world automotive case study (Schmitter, Ma, & Puschner, 2016) STPA-Sec was found to have significant limitations. The top-down approach was most applicable during the concept-phase of the SDLC but was insufficient on its own to satisfy all the co-assurance requirements.

*Security-Aware HAZOP.* SAHARA (Security-Aware Hazard and Risk Analysis Method) (Macher et al., 2015) is a HAZOP-like analysis for structured brainstorming with additional guidewords for security. A clear advantage of this method is that practitioners from both domains work together directly using shared concepts and terms. However, this method is time and resource-intensive due to the practicalities of organising the process and participants. It also assumes that everyone in the room has the right level of competency for the task.

*Fault Tree Analysis (FTA).* Integrated Fault and Attack Trees (Fovino, Masera, & De Cian, 2009) have been used to consider the interaction of malicious deliberate acts with random failures quantitatively. This analysis has been extended to include mitigations against some of the identified attack vectors (Kordy, Piètre-Cambacédès, & Schweitzer, 2014). The unambiguous semantics of using methods based on fault trees to represent both faults and threats has many benefits such as enabling practitioner to better understand some of the goals of the attacker. These methods suffer from similar limitations to FTA, where it is difficult to model dependency. This may lead to misidentification of attack paths which undermine the analysis.

*Dependability Analysis.* Dependability Deviation Analysis (DDA) is an analysis method used to identify potential failure conditions from the perspective of each quality attribute (Despotou, Alexander, & Kelly, 2009). DDA gives a multi-attribute perspective of the on the bow-tie analysis concept and thus provides a methodical way of identifying the links between safety and security failure conditions through the use of guidewords. Case studies of this methodology have been effective for complex systems (Despotou, 2007). The limitations of DDA include an over reliance on the participating practitioners to be know the impact of effects; in addition it is unclear how new results might be included during operation.

*Architectural Method.* The Architectural Trade-Off Analysis Method (ATAM) (Kazman et al., 1998) is a human-centric process for identifying risks early in the SDLC. It requires the software architects designing the system to gather and establish how a particular architecture satisfies given quality goals, and how the attributes trade off against each other. Typically, this process takes place over four days (Medvidovic & Taylor, 2010). This method is resource intensive and is usually most applicable during the design stage.

The last two methods are qualitatively different in their objectives to those preceding them, however they demonstrate the diversity of solutions available to this this problem. These analyses present a first step to integrated assurance. As briefly shown through the limitations of each of the methods, there remain several open problems that need to be resolved. In particular, it is unclear how to incorporate new security threat intelligence during the operational phase of the system without re-evaluating the entire system. This may not be possible, especially in light of the fact that several major security patches take place over a shorter period of time than it would take to perform the analyses.

**Risk Evaluation.** The risk aspect of the technical challenge is not independent from the analyses presented in the previous section. It is arguably the most difficult aspect of the technical challenge, therefore warrants its own discussion. The safety-security risk evaluation problem is how to measure, analyse, propagate and reason about risk. Large, complex systems increase the amount of uncertainty about system behaviour therefore making it difficult to accurately reason about risk especially using traditional causal models. In response to this problem there have been attempts in research and industry to create resources to understand and evaluate risk. Resources that include international cyber security incident reporting and monitoring (Johnson, 2015), frameworks to analyse sources, types, targets and motivations of attacks (Kshetri, 2005), and methods for evaluating damage from cyber attacks (Lala & Panda, 2001; Kundur et al., 2010) especially where they are linked to physical attacks. The following sections outline some of the key contributors to the risk evaluation challenge:

*Definition of Risk.* There is currently no widely accepted cross-domain definition of risk for safety and security. There are some conceptual models that include the two attributes (Firesmith, 2003), however these are insufficient to tackle the issue of risk propagation. Where safety risk is often characterized by severity and likelihood, security risk is characterised by many more factors such as impact and motivation. It is also more difficult to make a likelihood estimation for threats.

*Quantitative Risk Measure.* Researchers have attempted to use probability as a measure (Aven, 2007) and evaluate risk with a variation of probability risk assessment (Taylor, Krings, & Alves-Foss, 2002). However, the uncertainty in estimating risks precluded having a single, meaningful *quantitative* measure. Instead of being used as a direct measure, probability and likelihood can be used effectively to indicate the amount of confidence required for the assurance of a system, or sensitivity analysis. For example, opportunity and access can be used as a predictive indicator for likelihood of attack and managed according to the desired assurance level.

*Qaulitative Risk Measure.* There exist alternative *qualitative* measures for risk that have been widely used such as Common Criteria evaluation assurance levels (ISO/IEC, 2017) for security, and development assurance levels for safety (RTCA, 2012). These have proved useful when reasoning about individual attributes within specific domains but there has been no widely adopted or sophisticated integrated measure. It is important to note that a one-size-fits-all measurement that acts as magic bullet in unifying safety and security risk is not an adequate solution. Too much important information about uncertainty is discarded when these kinds of methods are adopted rendering them unfit for the purpose of accurately reasoning about risk. Instead, what is needed is a more nuanced way to reason about risk and track uncertainty.

*Risk Communication.* The communication of risk is related to the quantitative *vs* qualitative question. The lack of standardised models across domains leads to misunderstandings, lost information and asynchronous duplicate processes. Some research has been done into combining safety and security processes (Kriaa, Pietre-Cambacedes, Bouissou, & Halgand, 2015), argumentation approaches (Lautieri, Cooper, & Jackson, 2005), and controlled vocabularies for safety assurance (Attwood, Kelly, & Conmy, 2014). This work has predominantly been with just one of the attributes as the focus *e.g.* security-informed safety. In addition, many of the techniques have not shown adequate consideration to how teams currently work.

*Risk Representation.* Part of the communication problem is that it is unclear what constitutes a joint model or representation of risk. Both domains are over-reliant on expert knowledge which is often represented as text-based documents that are difficult to parse and update when change needs to be incorporated. Communication of expert knowledge is often ad-hoc or rigidly prescribed with little flexibility, such as with some of the technical analysis methods discussed earlier. The problem is further compounded by the lack of a shared language and terminology between teams, and lack of synchronised development techniques during the system lifecycle. As a result, with time, analysis models diverge and the link between safety and security becomes increasingly obscured. The trade-off therefore is unclear and a whole systems approach is almost impossible because the relevant information is provided long after the engineering decisions it would have influenced.

*Evolving Threat.* This aspect of the risk challenge is related to the increased cyber-security threat from activist, criminal and state-sponsored groups which are organised, have many resources, are highly motivated and can stage sophisticated attacks (Symantec, 2018). These attackers are able to exploit the increased number of attack vectors which result from greater system complexity *e.g.* increased zero-day vulnerabilities, as well as tried and tested methods *e.g.* spear phishing. Cyberterrorism remains poorly understood (Kenney, 2015) but still poses a unique and urgent threat to critical national infrastructure and SCS as it allows for greater damage to be done than using traditional weapons. Despite the abundance of work in this area there is still no consensus as to what the threats are or their impact. What is needed, therefore, is a way to reason about cyber risk that allows system development to progress without ignoring uncertainty and losing information that might be resolved at a later stage or with new technology or increased resources.

In addition to aspects of the technical challenge already mentioned, there has research which recognises some of the subtle interactions between safety and security (Lautieri, Cooper, & Jackson, 2005; Amorim et al., 2015). There has also been work done to reconcile safety-critical and high security functional requirements (Tomlinson et al., 2015), extend safety-security workflow tools (Schmittner, Althammer, & Gruber, 2015), combine safety and security in industrial control systems (Kriaa et al., 2015), extend the concept of assurance cases to security (Finnegan & McCaffery, 2014), and create complementary standards and codes of practice. However, there are no widely applied solutions for how to synchronously develop safety and security arguments during the system lifecycle, what information to share and how or when to share it. What is missing still is a fundamental philosophy, unifying language and standard set of practices for engineers to use during the system development. The next section discusses some of the socio-technical problems that arise due to this deficit.

**Socio-Technical Barriers to Co-Assurance**

In the previous section the technical difficulties of combing safety and security were discussed. These aspects are extensively covered in the literature, however in real-world systems they do not appear in isolation. Instead, they are part of an overall engineering process that is subject to drivers other than the technical. Therefore, no sustainable solution will be implemented without also addressing the socio-technical aspects of the challenge. The following discussion is not meant to be exhaustive, but it does provide an illustrative set of key areas that any solution would need to be address:

**Trade-Off.** Unlike other system quality attributes *e.g.* reliability, availability, maintainability, *etc.*, security poses a unique challenge to safety as it is not only a question of architectural and design trade-off. There exist more subtle ways in which arguments for safety are undermined and undercut by security threats (Bloomfield, Netkachova, & Stroud, 2013). This subtle interplay is not yet fully understood, and has not been fully addressed in current research. It can be thought of on different levels of abstraction:

*Conceptual Trade-Off.* Safety engineering has been established for more than 70 years and the conceptual framework that it works within is fairly mature. Techniques and language are fairly well established even if there is some debate within the domain. This, in addition to the fact that safety often takes precedence during the development of SCS, leads to an oversimplification of security assurance that lacks sufficient appreciation of what makes security risk reduction difficult. It is not enough to simply apply extant techniques from safety. Table 1 shows a few differences in philosophies that would affect the engineering and assurance processes. A fundamental difference is that safety is often non-negotiable and what is meant by 'harm' is fairly clear. Security 'harm', on the other hand, is less clear and dependent on the perspective. It is often about committing risk reduction resources proportionally to threats. It is much more difficult to assess whether a security goal has been attained.

Table 1 — **Key Differences in Philosophies**

| Safety | Security |
|---|---|
| - predominantly values domain openness, collaboration, transparency<br>- assumes accidents happen as a result of random and unintentional failures<br>- assumes a benevolent operator | - security-through obscurity and information hiding are valid controls<br>- assumes a space of adversarial competition with fast-evolving threats from intelligent attackers who have potentially infinite attack vectors |

*Organisational Trade-Off Considerations.* Within organisations the safety and security communities are often physically separate. The practitioners in each domain tend to specialise in very detailed but often disparate knowledge. This becomes problematic when conflicting concerns need to be resolved. In addition to the physical separation, there is often a mismatch in the number of engineers on projects. Safety teams are often well-established and relatively large compared to the small security teams which have fewer practitioners with the right competency level (Bird, 2017; Ullrich, 2016). This presents many practical problems, such as security engineers may not be able to attend as many meetings as the safety team which would greatly affect some of the analysis methods described in the previous chapter.

*Trade-Off Considerations for Individuals.* Understanding the implications of trade-off is difficult because it requires an understanding of complex interactions, and needs a practitioner to access higher creative cognitive functions. It can be argued that it is unlikely that a practitioner from a single domain would have the oversight and authority to make judgement calls about impact in another domain on their own. If this were attempted then any results would likely be subject to several biases, such as confirmation bias and the Dunning-Kruger effect (Dunning, 2011) which can lead to overconfidence in a safety argument because of a lack of understanding of security, for example. A large part of understanding security relates to the attacker and their motivations. The next section explores some of the difficulties introduced by having an adversary.

**Adversarial Nature of Security.** Previous sections have discussed security being in an adversarial space. The presence of an intelligent attacker means that conflict with safety cannot only be resolved through trade-off alone. In many cases the adversarial nature of attackers causes a relationship where security is inversely proportional to safety. For example, safety certification requires a transparent argument that a system will perform its intended function in a safe way. This argument structure provides potential attackers with a clear blueprint of system weaknesses and attack vectors. It is often the motive, not means that explains the absence of an attack. The implications are that there is a greater need to understand the security argument of a system and the reasons to have confidence in it in order to better understand security risk.

**Proportionality**. The concept of proportionality is not a new one in system assurance. It defines the view that the measures taken and resources allocated to control risk and must be proportional to the magnitude of the risk itself (Zakaszweska, 2016). There are several aspects of the risk management process that proportionality affects, namely the amount of dedicated process, how much time is afforded to risk management, the competence that is required, the detail of evidence and the level of assurance. Existing technical solutions to the safety-security challenge do not seem to consider all proportionality aspects. With reference to competence, it is often assumed that the practitioners performing the analysis are suitably qualified, however for security, one of the top challenges consistently identified is lack of skills (Bird, 2017; Ullrich, 2016).

The aim of providing this very brief, but detailed discussion of the key problems identified for the safety-security challenge is to draw attention to the challenges and gaps in knowledge that still exist. This is important when creating a new solution, in order to avoid being subject to the same limitations.

## Safety-Security Assurance Framework (SSAF)

Having enumerated the existing techniques to solve the safety-security challenge, and discussed the socio-technical issues surrounding the problem, in this section a candidate solution that attempts to address some of these problems is proposed. This is the Safety-Security Assurance Framework.

### Independent Co-Assurance as a Solution

The many reasons why safety and security assurance cannot remain predominantly independent have already been discussed. So, too, have the reasons why the attributes cannot be simply unified in one assurance process. A better candidate solution is one that lies between the two extremes, that allows for independently running assurance activities, but has synchronisation points where risk information is propagated. This model of activity is what is being defined as independent co-assurance. In order for this approach to be successful and effective, it requires a common base. Thus, to achieve this common understanding, the safety assurance principles previously identified from standards (Hawkins, Habli, & Kelly, 2013) have been applied to security with the following outcome:

1. Software security requirements shall be defined to address the software contribution to system vulnerabilities.
2. The intent of the software security requirements shall be maintained throughout requirements decomposition.
3. Software security requirements shall be satisfied.
4. Vulnerabilities introduced by software behaviour shall identified be mitigated.
4+1. The confidence established in addressing the software security principles shall be commensurate to the contribution of the software to system risk.

Whilst this is seemingly an exercise in renaming the principles from safety to security, the implications are a greater. It creates a common assurance argument structure which can be used as the basis for communication during indpedendent assurance activities. It changes co-assurance activities from a process of integrating safety and security in very specific ways at very specific times, to a question of activity synchronisation that allows for greater flexibility. In addition, this solution uses the model-based system engineering paradigm (Wymore, 1993) to integrate safety and security assurance activities with each other, and with the SDLC. It functions by allowing safety and security teams to work separately, but defines points at which they must share information to produce an integrated assurance case. This is a highly innovative solution because it aims to keep the benefits of working in specialised teams whilst still producing an integrated assurance argument for the system. This principle-based approach ultimately is a lot more suited to real-world application where assurance of the attributes is unlikely to be at the same rate or by the same team.

**What is SSAF? What does it consist of?**

The solution, as illustrated in Figure 1, consists of:

**Process**
- steps to develop an integrated assurance argument structure
- points of communication during system development
- a method of risk propagation and management
- steps to configure or restrict information sharing

**Models**
- a meta-model for safety and security assurance artefacts
- common argument patterns for safety and security
- examples of links between the artefacts generated form particular methodologies

**Language**
- ontology of terms and concepts
- a method for standardising language and terminology used during assurance

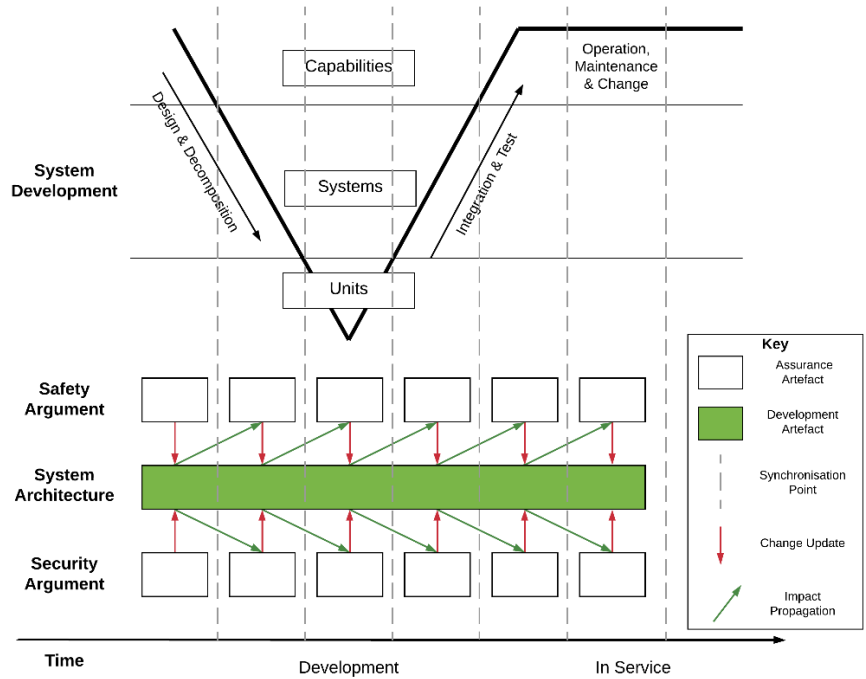

**Figure 1— Proposed Safety-Security Assurance Framework (SSAF)**

**How will SSAF be implemented?**

SSAF makes use of the UML-based standard for structured assurance cases: Structured Assurance Case Metamodel (SACM) (OMG, 2018). This will allow models to be built that include detailed information about artefacts generated from specific activities by participants. These artefacts will also relate to particular claims. Relations between the artefacts from each of the assurance activities will then be created, and these would be the vehicles for impact propagation *e.g.* when a vulnerability artefact is changed, any hazard artefacts related to it will be updated.

**Synchronisation**

This candidate solution provides a process for separate safety and security assurance and expertise, but facilitates synchronised co-evolution through the SDLC. This framework will allow for controlled information sharing and directly addresses several aspects of the safety-security challenge. It allows for better communication using the same language and terminology. This will limit the separate analyses from diverging from each other. The traceable link through the lifetime of the system is maintained in this way.

**Attribute Co-existence**

The SSAF aims to go beyond simple high-level issue flagging or updates on measures. It will provide a way to reason about the subtle ways in which claims interact with each other through their associated artefacts. It is an improvement on existing methods because it makes it necessary to articulate claims in a standardised form. This, in turn, allows practitioners to evaluate risk and impact at a deeper level which does not obscure information. The solution also formalises how system and safety-security assurance models relate to each other, this creates the potential for partial automation through model-based practices that are already established.

## Expected Outcomes from SSAF

The primary outcome of the SSAF is that the safety and security arguments are made explicit and linked to the system model so that justifications and impact are clearer. These argument structures, represented as models, will also be used as the primary source of information for certification and accreditation. Over time it is expected that patterns for the structures will be derived.

The advantages of this solution include, but are not restricted to, that it will harness the emergent benefits and capabilities of new technology without counteractively restricting activities. The impact, trade-offs and uncertainty of safety-security interactions will be more traceable. The solution will enable better arguments to be formed, and enable better decisions regarding the system because the uncertainty related to an argument are presented in a transparent way. It is not a one-size-fits-all representation of risk that is blindly applied to all situations. Rather, it enables risk measures to be applied to safety and security arguments with a degree of confidence which can be revisited at a later stage, this allows for more sophisticated reasoning.

## Conclusions

This paper has discussed some of the major challenges and gaps in knowledge related to safety and security assurance of large, complex systems. These gaps related to the differences between safety and security communities, how to represent and reason about risk, and how arguments can be represented as models. The safety-security assurance framework (SSAF), was presented as a candidate solution to these challenges which aims to create a process for synchronising the independent assurance of safety and security, and create a more sophisticated and nuanced way to reason about impact. The SSAF has the potential to positively change the way safety and security communities interact with each other, especially when developing large, complex systems where uncertainty is high. As a result it is possible that the systems become safer and more secure as a result of this framework.

## Acknowledgement

This project is funded by the UK's Engineering and Physical Sciences Research Council through an Industrial Cooperative Award in Science & Technology (EPSRC iCASE) studentship, in partnership with BAE Systems and the University of York.